\documentclass[runningheads]{llncs}
\usepackage[T1]{fontenc}
\usepackage{graphicx}
\usepackage{orcidlink}
\usepackage{amsmath}
\usepackage{amsfonts}
\usepackage{tikz}
\usepackage{hyperref}
\usepackage{cleveref}
\usepackage{color}

\usepackage{subcaption}
\usepackage{orcidlink}
 \usepackage[misc]{ifsym}

\newcommand{\ap}{\mathit{AP}}

\newcommand{\calT}{\mathcal{T}}

\providecommand{\ltlN}{\operatorname{%
		\protect\tikz[baseline]{
			\draw[line width=.12ex]
			(0,.6ex) circle (.8ex);
}}}{}

{}

\providecommand{\ltlG}{\operatorname{%
		\protect\tikz[baseline]{
			\draw[line width=.12ex,line join=round]
			(0ex,-.2ex) -- (0ex,1.3ex) -- (1.5ex,1.3ex) -- (1.5ex,.-.2ex) -- cycle;
}}}{}

\DeclareMathOperator{\ltlU}{\mathcal{U}}

\newcommand\xqed[1]{%
	\leavevmode\unskip\penalty9999 \hbox{}\nobreak\hfill
	\quad\hbox{#1}}
\newcommand\demo{\xqed{$\triangle$}}

\newcommand{\tool}{\texttt{HyGaViz}}

\newcommand{\ldot}{\mathpunct{.}}

\newif\ifbadgeavailable\newif\ifbadgefunctional\newif\ifbadgereusable
\badgeavailabletrue
\badgefunctionaltrue
\badgereusablefalse
\RequirePackage{graphicx}
\usepackage[firstpageonly=true,angle=0,vpos=.147\paperheight,hpos=.393\linewidth,vanchor=t,hanchor=l]{draftwatermark}
\SetWatermarkText{%
	\raisebox{-3.36cm}
	{\ifbadgeavailable\includegraphics[width=11mm]{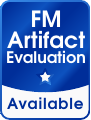}\hspace{.815\linewidth}\else\hspace{.905\linewidth}\fi%
		\ifbadgefunctional\includegraphics[width=11mm]{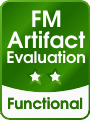}\else\ifbadgereusable\includegraphics[width=11mm]{FM_2024_AE_reusable}\fi\fi}}

\begin{document}
\title{Visualizing Game-Based Certificates for Hyperproperty Verification}
\titlerunning{Visualizing Game-Based Certificates for Hyperproperty Verification}
%

\author{Raven Beutner\orcidlink{0000-0001-6234-5651} \and
	Bernd Finkbeiner\orcidlink{0000-0002-4280-8441} \and
	Angelina Göbl\orcidlink{0009-0009-2331-4049}}
\authorrunning{R.~Beutner, B.~Finkbeiner, A.~Göbl}
%

\institute{CISPA Helmholtz Center for Information Security, \\ Saarbrücken, Germany \\
\email{\{raven.beutner,finkbeiner,angelina.goebl\}@cispa.de}
	}
\maketitle              
\begin{abstract}
	Hyperproperties relate multiple executions of a system and are commonly used to specify security and information-flow policies. 
	While many verification approaches for hyperproperties exist, providing a convincing \emph{certificate} that the system satisfies a given property is still a major challenge.
	In this paper, we propose \emph{strategies} as a suitable form of certificate for hyperproperties specified in a fragment of the temporal logic HyperLTL.
	Concretely, we interpret the verification of a HyperLTL property as a game between universal and existential quantification, allowing us to leverage strategies for the existential quantifiers as certificates. 
	We present \tool{}, a browser-based visualization tool that lets users interactively explore an (automatically synthesized) witness strategy by taking control over universally quantified executions. 
\end{abstract}

\section{Introduction}

Hyperproperties \cite{ClarksonS10} relate multiple execution traces of a system and occur frequently when reasoning about information flow \cite{ZdancewicM03,Rabe16}, robustness \cite{BiewerDFGHHM22,ChaudhuriGL12}, independence \cite{BartocciHNC23}, knowledge \cite{BozzelliMP15,BeutnerFFM23}, and causality \cite{CoenenFFHMS22,FinkbeinerS23}.
A popular logic for specifying temporal hyperproperties is HyperLTL \cite{ClarksonFKMRS14}, an extension of LTL with explicit quantification over execution traces.
For example, we can use HyperLTL to express a simple non-interference property as follows:
\begin{align}\label{eq:gni}
	\forall \pi_1. \exists \pi_2 \ldot \ltlG (l_{\pi_1} \leftrightarrow l_{\pi_2}) \land \ltlG(o_{\pi_1} \leftrightarrow o_{\pi_2}) \land \ltlG (\neg h_{\pi_2}) \tag{$\varphi_\mathit{NI}$}
\end{align} 
Informally, this property -- called non-inference \cite{McLean94} -- requires that any possible observation made via the low-security input (modeled via atomic proposition $l$) and output ($o$) is compatible with a fixed ``dummy'' sequence of high-security inputs ($h$) \cite{McLean94}.
Concretely, \ref{eq:gni} states that for \emph{any} execution $\pi_1$, \emph{some} execution $\pi_2$ combines the low-security observations of $\pi_1$ with fixed dummy values for $h$; here, we require that $h$ is constantly set to false, i.e., $\ltlG (\neg h_{\pi_2})$ (cf.~\cite{Finkbeiner23}).

\paragraph{Verification and Certificates.}

In recent years, many verification techniques for temporal hyperproperties (expressed, e.g., in HyperLTL) have been developed \cite{ClarksonFKMRS14,FinkbeinerRS15,BeutnerF23,HsuSB21,BartheDR11,BeutnerF24,Rabe16}. 
However, while \emph{checking} if a given system satisfies a HyperLTL property is important, an often equally critical aspect is to convince the user of this satisfaction using explainable \emph{certificates}.
For trace properties -- specified, e.g., in LTL -- user-understandable certificates for positive and negative verification results have been explored extensively \cite{KasenbergTS20,GriggioRT18,BeerBCOT09,BeschastnikhLXW20,BoltonB10,GroceKL04}.
Likewise, for alternation-free HyperLTL formulas  (i.e., formulas that use a single type of quantifier), known techniques for LTL apply \cite{HorakCMHFMDFD22}. 
In contrast, generating explainable certificates for the satisfaction of alternating properties like \ref{eq:gni} is more complex. 
For example, \ref{eq:gni} states that for any trace $\pi_1$, there exists some matching execution $\pi_2$.
A certificate must thus implicitly define a mapping that, given a concrete choice for $\pi_1$, produces a witness trace $\pi_2$.
Defining and understanding such a mapping can be complex, even for simple systems with few states.

\paragraph{Strategies as Certificates.}

In this paper, we propose \emph{strategies} as certificates for the satisfaction of $\forall^*\exists^*$ HyperLTL formulas (i.e., formulas where an arbitrary number of universal quantifiers is followed by an arbitrary number of existential quantifiers; e.g., \ref{eq:gni}). 
To accomplish this, we take a game-based verification perspective \cite{CoenenFST19,BeutnerF22b}. 
The key idea is to interpret the verification of a $\forall \pi_1. \exists \pi_2\ldot \psi$ formula (where $\psi$ is the LTL body) as a game between universal and existential quantification.
The $\forall$-player controls the universally quantified trace by moving through the system (thereby producing a trace $\pi_1$), and the $\exists$-player reacts with moves in a separate copy of the system (thereby producing a trace $\pi_2$). 
Any strategy for the $\exists$-player that ensures that $\pi_1$ and $\pi_2$, together, satisfy $\psi$, implies that the formula is satisfied on the given system. 
We can think of a winning strategy as a step-wise Skolem function that, for every trace $\pi_1$, iteratively constructs a witnessing trace $\pi_2$.

\paragraph{Visualizing Strategies.}

In this paper, we introduce \tool{}, a verification and visualization tool for strategies in the context of HyperLTL verification. 
In \tool{}, the user can input (possibly identical) finite-state transition systems and a HyperLTL formula $\varphi$. 
\tool{} then automatically attempts to synthesize a strategy that witnesses the satisfaction of $\varphi$. 
If a strategy exists,  \tool{} displays it to the user. 
Our key insight is that we can let the user explore the strategy interactively by taking control of universally quantified traces. 
That is, instead of displaying the strategy in its entirety (e.g., as a table or decision diagram), we let the user play a game.
In each step of the game, the user decides on a successor state for each universally quantified system (i.e., the user takes the role of the $\forall$-player), and \tool{} automatically updates the states of all existentially quantified systems (i.e., \tool{} plays the role of the $\exists$-player).

\begin{figure}[!t]
	\begin{subfigure}{1.0\linewidth}
		\centering
		\includegraphics[width=1.0\linewidth]{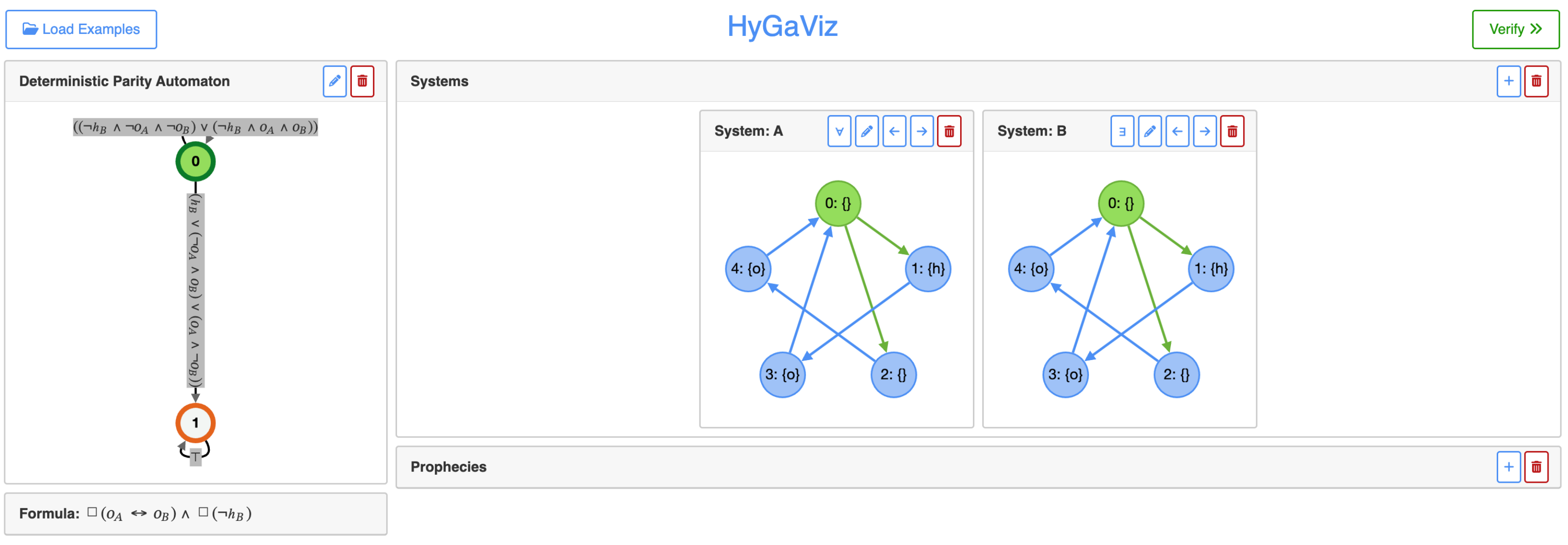}
		\vspace{-3mm}
		\subcaption{}\label{fig:overview-start}
	\end{subfigure}\\[3mm]
	\begin{subfigure}{1.0\linewidth}
		\centering
		\includegraphics[width=1.0\linewidth]{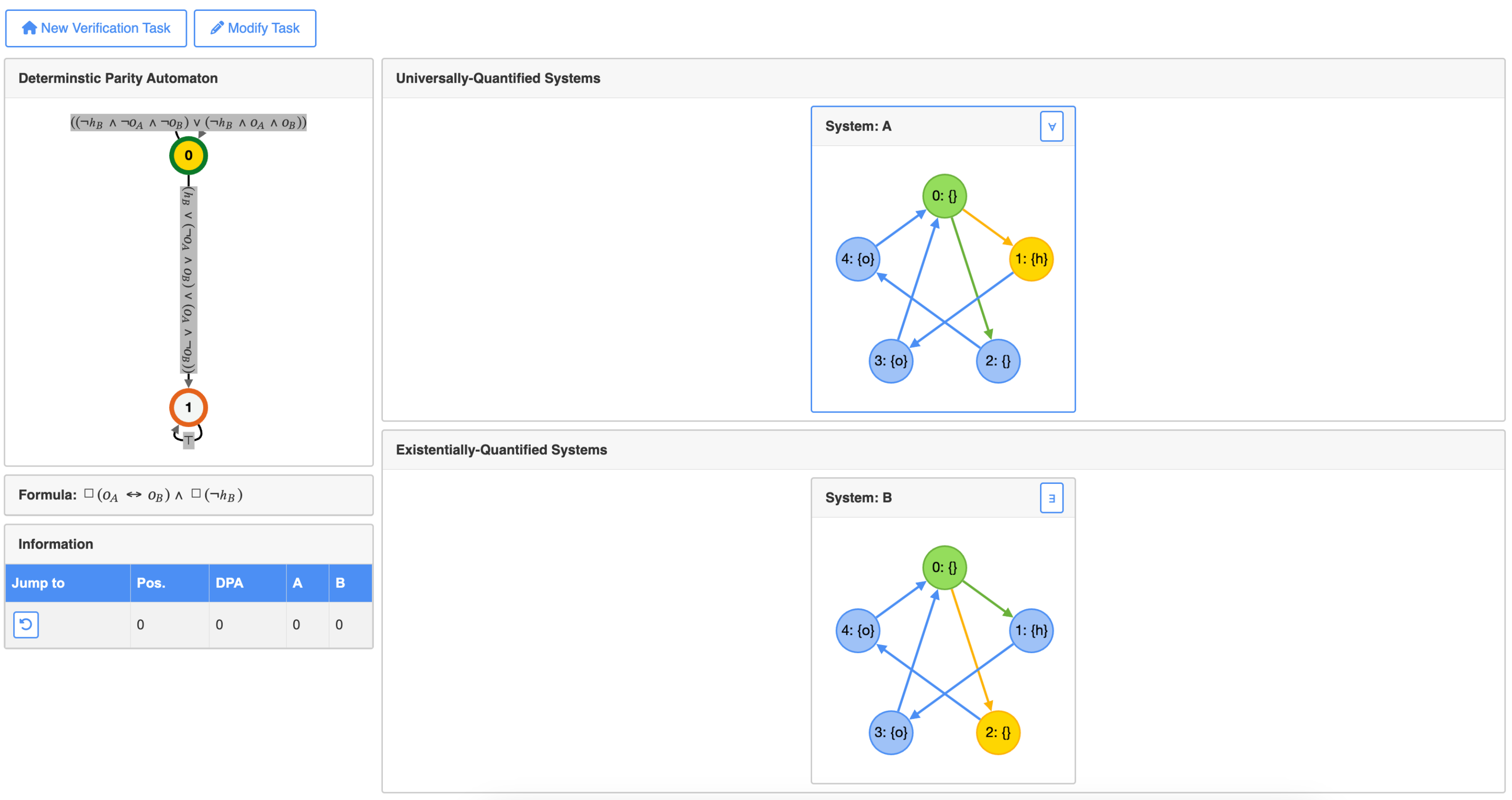}
		\subcaption{}\label{fig:overview-verify}
	\end{subfigure}
	\caption{Screenshots of \tool{}.}\label{fig:overview}
\end{figure}

\begin{example}
	We consider a simple verification instance in \Cref{fig:overview}.
	On \tool{}'s initial page (\Cref{fig:overview-start}), we create two (in this case, equal) transition systems (labeled $A, B$) over atomic propositions (APs) $o$ and $h$, depicted in the top right.  
	In each system, each state is identified by a natural number and lists all APs that hold in the given state.
	From initial state $0$, the system can branch on AP $h$ (states $1$ and $2$), but, in either case, AP $o$ is set in the next step (states 3 and 4).
	We want to verify \ref{eq:gni}, which -- due to the absence of low-security input $l$ -- simplifies to $\forall A. \exists B\ldot \protect\ltlG (o_A \leftrightarrow o_B) \land \ltlG (\neg h_B)$.
	Note how, in \tool{}, the quantifier prefix is determined implicitly by the order and quantifier type of the systems, and the LTL body is displayed on the bottom left.
	The user can change the systems, the quantification type, the name, and the order of the systems using the buttons above each system. 
	Upon entering the LTL formula,  \tool{} automatically displays a deterministic automaton for the property (top left). 
	After clicking the \emph{Verify} button (top right), the user is directed to the strategy simulation page (depicted in \Cref{fig:overview-verify}).
	During the simulation, \tool{} displays the current state of the automaton and the system state for $A$ and $B$ (in green) and lets the user control the state of (the universally quantified) system $A$.
	By hovering over the successor state of system $A$, \tool{} highlights the next state for system $B$ (in yellow). 
	In this instance, systems $A$ and $B$ are both in state $0$.
	When the user moves system $A$ to state $1$, \tool{} reacts by moving system $B$ to state $2$ (as it has to ensure $\ltlG (\neg h_B)$).
	By clicking on a successor state for $A$, the user locks the choice, and the game progresses to the next round.
	 \demo
\end{example}

\subsubsection*{Related Work.}

\texttt{HyperVis} \cite{HorakCMHFMDFD22} is a tool for the visualization of counterexample traces for alternation-free $\forall^k$ formulas.
Notably, a counterexample to a $\forall^k$ property is a concrete list of $k$ traces, so visualization is possible by highlighting the relevant parts of the traces, potentially using causality-based techniques \cite{CoenenDFFHHMS22}. 
Our visualization for properties involving quantifier alternations is rooted in the game-based verification approach for HyperLTL \cite{CoenenFST19,BeutnerF22b}, which becomes complete when adding prophecies \cite{BeutnerF22b} (see \Cref{sub:proph}).
To the best of our knowledge, we are the first to propose a principled approach to generate and visualize user-understandable certificates for alternating hyperproperties.

 \section{HyperLTL, Game-Based Verification, and Prophecies}
 
We fix a finite set of atomic propositions $\ap$.
A transition system (TS) is a tuple $\calT = (S, s_\mathit{init}, \kappa, L)$, where $S$ is a finite set of states, $s_\mathit{init} \in S$ is an initial state, $\kappa : S \to (2^S \setminus \{\emptyset\})$ is a transition function, and $L: S \to 2^\ap$ is a state labeling. 
HyperLTL formulas are generate by the following grammar 
\begin{align*}
	\psi &:= a_\pi \mid \psi \land \psi \mid \neg \psi \mid \ltlN \psi  \mid \psi \ltlU \psi \quad\quad\quad 	\varphi :=\forall \pi \ldot \varphi \mid \exists \pi \ldot \varphi \mid \psi
\end{align*}
where $a \in \ap$ is an atomic proposition, and $\pi$ is a trace variable. 
In a HyperLTL formula, we can quantify over traces in a system (bound to some trace variable), and then evaluate an LTL formula on the resulting traces. 
In the LTL body, formula $a_\pi$ expresses that AP $a$ should hold in the current step on the trace bound to trace variable $\pi$.
See \cite{Finkbeiner23} for details.

 \subsection{Game-Based Verification}

\tool{}'s verification certificates are rooted in a game-based verification method \cite{BeutnerF22b}.
Given a $\forall^*\exists^*$ HyperLTL formula $\forall \pi_1\ldots \forall \pi_k\ldot \exists \pi_{k+1}\ldots \exists \pi_{k+l}\ldot \psi$, we view verification as a game between the $\forall$-player (controlling traces $\pi_1, \ldots, \pi_{k}$) and the $\exists$-player (controlling traces $\pi_{k+1} \ldots, \pi_{k+l}$).
Each state of the game has the form $\langle s_1, \ldots, s_{k+l}, q\rangle$, where $s_1, \ldots, s_{k+l} \in S$ are system states (representing the current state of $\pi_1, \ldots, \pi_{k+l}$, respectively), and $q$ is the state of a deterministic parity automaton (DPA) that tracks the acceptance of the LTL body $\psi$. 
When the game is in state $\langle s_1, \ldots, s_{k+l}, q \rangle$, the $\forall$-player first fixes successor states $s_1', \ldots, s_k'$ for $\pi_1, \ldots, \pi_k$ (such that $s_i' \in \kappa(s_i)$ for all $1 \leq i \leq k$); the $\exists$-player responds by selecting successor states $s_{k+1}', \ldots, s_{k+l}'$ for $\pi_{k+1}, \ldots, \pi_{k+l}$; and the game repeats from state $\langle s_1', \ldots, s_{k+l}', q'\rangle$ (where $q'$ is the updated DPA state).

\paragraph{Visualizing Game-Based Verification.}

In \tool{}, the user can create a verification scenario by manually creating finite-state transition systems and a HyperLTL formula; see \Cref{fig:overview-start}.
Note how the quantification prefix is determined implicitly by the order of the systems. 
In particular, the traces are resolved on individual (potentially different) transition systems. 
During simulation (cf.~the example in \Cref{fig:overview-verify}), we visualize a game state $\langle s_1, \ldots, s_{k+l}, q \rangle$ by marking the current state of each system -- separated into user-controlled (universally quantified) systems (top right) and strategy-controlled (existentially quantified) ones (bottom right) -- and display the current state of the DPA (top left).
The user takes the role of the $\forall$-player and, in each step, determines successor states for all universally quantified systems. 
Once successor states for all universally quantified systems are confirmed, \tool{} automatically updates existentially quantified systems (and the DPA state) based on the internally computed strategy, and the game continues to the next stage.
Moreover, \tool{} highlights the next states when the user \emph{hovers} over possible successor states for the universally quantified systems (once successor states for all but one universally quantified system are confirmed). 
Using the information tab in the bottom left, the user can jump to previous game states and explore the reaction of the strategy to different choices for the universally quantified systems.

 \begin{figure}[!t]
 	\begin{subfigure}{1.0\linewidth}
 		\centering
 		\includegraphics[width=1.0\linewidth]{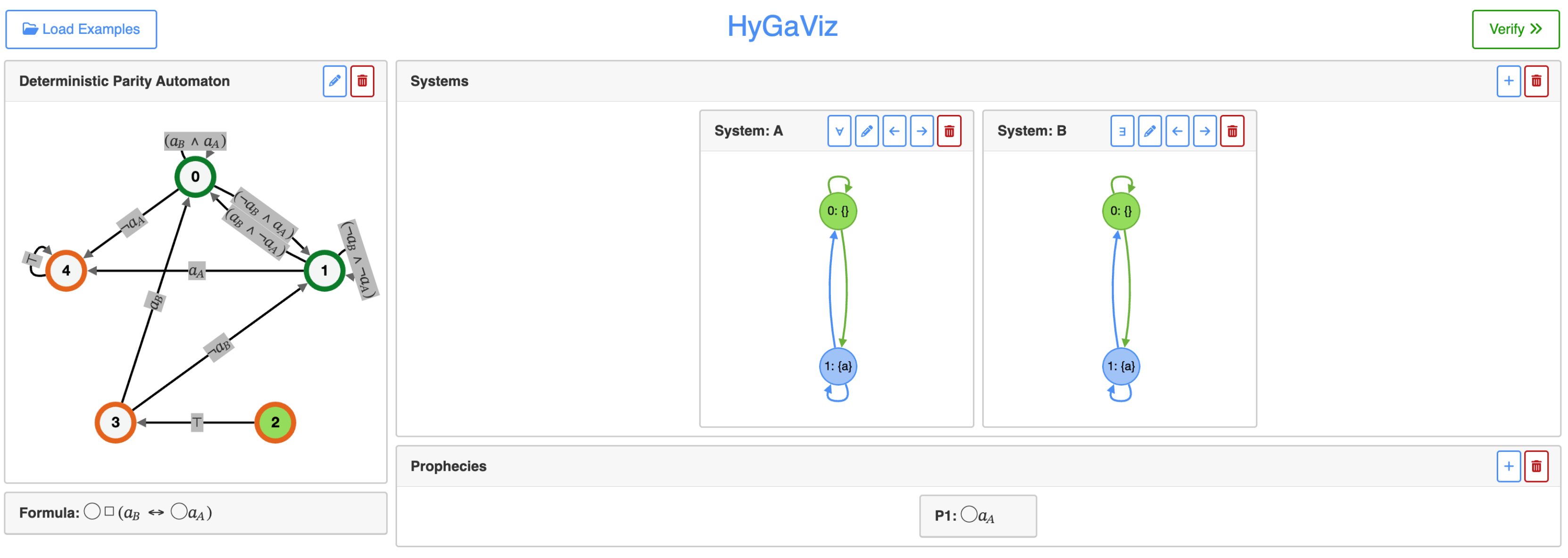}
 		\subcaption{}\label{fig:proph-start}
 	\end{subfigure}\\[4mm]
	\begin{subfigure}{0.5\linewidth}
		\centering
		\includegraphics[width=0.8\linewidth]{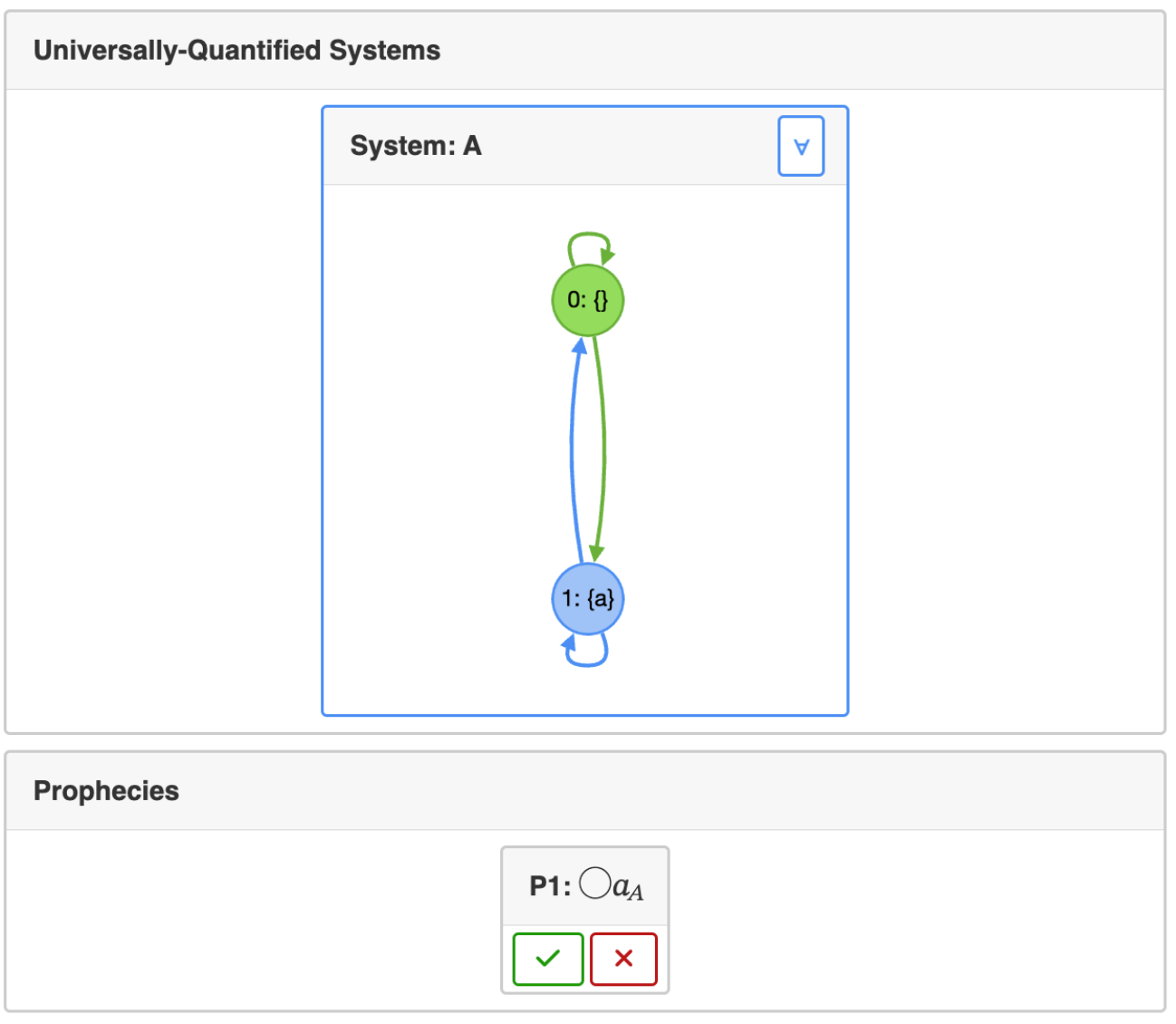}
		\subcaption{}\label{fig:proph-verify}
	\end{subfigure}%
	\begin{subfigure}{0.5\linewidth}
		\centering
		\includegraphics[width=0.9\linewidth]{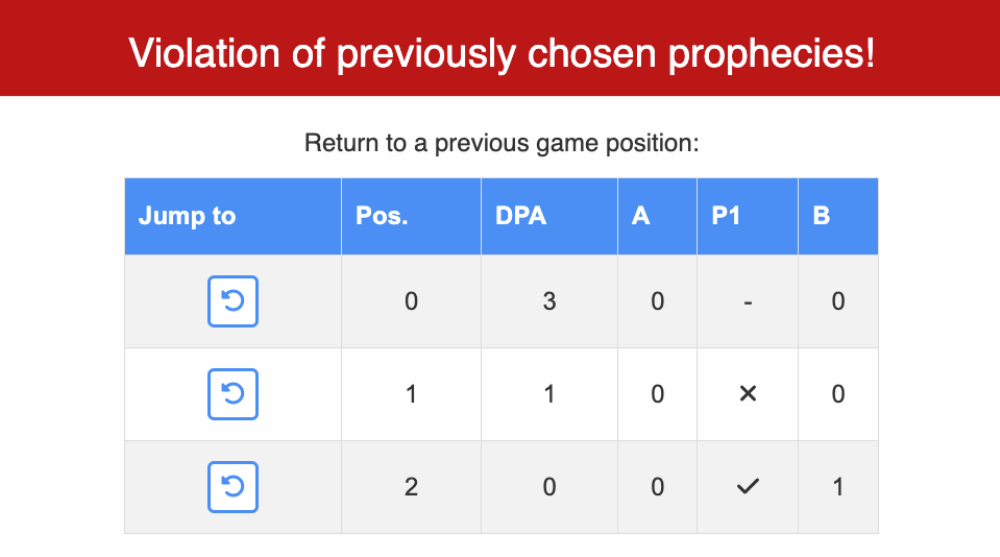}
		\vspace{4mm}
		\subcaption{}\label{fig:proph-vio}
	\end{subfigure}
	\caption{Screenshots of \tool{} when using prophecies.}\label{fig:proph}
\end{figure}

\subsection{Prophecies}\label{sub:proph}

In our game, the $\exists$-player only observes a finite prefix of the traces produced by the $\forall$-player (or, equivalently, the user of \tool{}) and is thus missing information about the future. 
We can counteract this by using \emph{prophecies} \cite{AbadiL91}, which are LTL formulas over trace variables $\pi_1, \ldots, \pi_k$ \cite{BeutnerF22b}. 
Given an LTL prophecy formula $\theta$, the $\forall$-player (i.e., the user) has to, in each step, decide if its future behavior (on $\pi_1, \ldots, \pi_k$) satisfies $\theta$. 
If the $\forall$-player decides that $\theta$ holds (resp.~does not hold), the $\exists$-player can play under the assumption that the \emph{future} behavior of the $\forall$-player satisfies (resp.~violates) $\theta$.
See \cite{BeutnerF22b} for details. 

\begin{example}
	We illustrate prophecies with the example in \Cref{fig:proph}.
	The two systems $A$ and $B$ in \Cref{fig:proph-start} generate all traces over AP $a$, and the HyperLTL formula $\forall A\ldot \exists B\ldot \ltlN \ltlG (a_B \leftrightarrow \ltlN a_A)$ requires that trace $B$ \emph{predicts} the future behavior of $A$. 
	Without prophecies, the $\exists$-player loses: No matter what successor state the $\exists$-player picks, the $\forall$-player can, in the next step, violate the prediction of the $\exists$-player. 
	\tool{} communicates the absence of a winning strategy if the user pushes the \emph{Verify} button. 
	Instead, the user can add the LTL prophecy $\ltlN a_A$ (cf.~\Cref{fig:proph-start}).
	During simulation, the user (who takes the role of the $\forall$-player) has to, in each step, fix a successor state for system $A$ \emph{and} determine if prophecy $\ltlN a_A$ holds. 
	We depict an excerpt of the simulation page in \Cref{fig:proph-verify}.
	As expected, the strategy for the $\exists$-player (computed automatically by \tool{}) can use the prophecy to win: 
	For example, if the user states that $\ltlN a_A$ holds (so the $\exists$-player can assume that $a$ hold in the next step in $A$), \tool{} moves system $B$ to state $1$.
	If the user violates a previous prophecy decision -- e.g., by stating that prophecy $\ltlN a_A$ holds but, in the next step, moving system $A$ to state $0$ where AP $a$ does not hold -- \tool{} detects this violation and forces the user to restart from an earlier state of the game (\Cref{fig:proph-vio}). \demo
\end{example}

\section{HyGaViz: Tool Overview}

\tool{} consists of a backend verification engine written in \texttt{F\#}.
The backend uses \texttt{spot} \cite{Duret-LutzRCRAS22} to translate LTL formulas to DPAs and \texttt{oink} \cite{Dijk18} to synthesize a strategy for the $\exists$-player. 
We use a stateless \texttt{Node.js} \cite{TilkovV10} backend that communicates with the verification engine via \texttt{JSON}. 
\tool{}'s frontend is written in \texttt{JavaScript} and uses \texttt{Cytoscape.js} \cite{FranzLHDSB16} to render transition systems and automata.

\section{Conclusion}

We have proposed the first method to generate and visualize certificates for the satisfaction of $\forall^*\exists^*$ HyperLTL formulas. 
Our tool, \tool{}, allows users to \emph{interactively} explore the complex dependencies between multiple traces by challenging a strategy for existentially quantified traces.
Ultimately, \tool{} is a first step to foster trust in (and understanding of) verification results for complex alternating hyperproperties, as is needed to, e.g., certify information-flow policies like \ref{eq:gni}. 
For now, \tool{} can handle (small) finite state systems, which we visualize as directed graphs. 
The underlying strategy-centered approach also applies to larger (potentially infinite-state) systems represented symbolically \cite{BeutnerF22}. 
In future work, one could extend \tool{} to such systems by exploring different visualization approaches for larger systems \cite{MorenoMSB04,JerdingSB97,RajalaLKS08}. 

\subsubsection*{Data Availability.}

\tool{} is available at \cite{beutner_2024_12206584}.

\subsubsection*{Acknowledgments.}

This work was partially supported by the European Research Council (ERC) Grant HYPER (101055412) and by the German Research Foundation (DFG) as part of TRR 248 (389792660).

%
%
%
 \bibliographystyle{splncs04}
 \bibliography{references}

\end{document}